\documentclass[useAMS,usenatbib]{mn2e}
\usepackage{amssymb,latexsym,graphicx,natbib,times}
\pdfoutput=1
\newcommand{\mi}{\relax \ifmmode {\mu{\mbox m}}\else $\mu$m\fi}
\newcommand{\hii}{\relax \ifmmode {\mbox H\,{\scshape ii}}\else H\,{\scshape ii}\fi}
\newcommand{\sii}{\relax \ifmmode {\mbox S\,{\scshape ii}}\else S\,{\scshape ii}\fi}
\newcommand{\nii}{\relax \ifmmode {\mbox N\,{\scshape ii}}\else N\,{\scshape ii}\fi}
\newcommand{\neii}{\relax \ifmmode {\mbox Ne\,{\scshape ii}}\else Ne\,{\scshape ii}\fi}
\newcommand{\neiii}{\relax \ifmmode {\mbox Ne\,{\scshape iii}}\else Ne\,{\scshape iii}\fi}
\newcommand{\oiii}{\relax \ifmmode {\mbox O\,{\scshape iii}}\else O\,{\scshape iii}\fi}
\newcommand{\oii}{\relax \ifmmode {\mbox O\,{\scshape ii}}\else O\,{\scshape ii}\fi}
\newcommand{\oi}{\relax \ifmmode {\mbox O\,{\scshape i}}\else O\,{\scshape i}\fi}
\newcommand{\ha}{\relax \ifmmode {\mbox H}\alpha\else H$\alpha$\fi}

\newcommand{\hep}{\relax \ifmmode {\mbox H}\epsilon\else H$\epsilon$\fi}
\newcommand{\hdel}{\relax \ifmmode {\mbox H}\delta\else H$\delta$\fi}
\newcommand{\hgam}{\relax \ifmmode {\mbox H}\gamma\else H$\gamma$\fi}

\newcommand{\pa}{\relax \ifmmode {\mbox Pa}\alpha\else Pa$\alpha$\fi}
\newcommand{\hb}{\relax \ifmmode {\mbox H}\beta\else H$\beta$\fi}
\newcommand{\rdostres}{\relax \ifmmode {\,\mbox{R}}_{\rm 23}\else \,\mbox{R}$_{\rm 23}$\fi}
\newcommand{\ergs}{\relax \ifmmode {\,\mbox{erg\,s}}^{-1}\else \,\mbox{erg\,s}$^{-1}$\fi}
\newcommand{\me}{\relax \ifmmode {\,}^{-1}\else \,$^{-1}$\fi}

\newcommand{\msun}{\relax \ifmmode {\,\mbox{M}}_{\odot}\else \,\mbox{M}$_{\odot}$\fi}

\newcommand{\cmtres}{\relax \ifmmode {\,\mbox{cm}}^{-3}\else \,\mbox{cm}$^{-3}$\fi}
\newcommand{\cmdos}{\relax \ifmmode {\,\mbox{cm}}^{-2}\else \,\mbox{cm}$^{-2}$\fi}
\newcommand{\cmseis}{\relax \ifmmode {\,\mbox{cm}}^{-6}\else \,\mbox{cm}$^{-6}$\fi}
\newcommand{\hi}{\relax \ifmmode {\mbox H\,{\scshape i}}\else H\,{\scshape i}\fi}

\title[Leakage effect on Star Formation Rate estimates]{On how leakage can affect the Star Formation Rate estimation using \ha\ luminosity}

\author[M. Rela\~no et al.]{M. Rela\~no$^{1}$\thanks{E-mail:
mrelano@ugr.es}, R. C. Jr. Kennicutt$^{2}$, J. J. Eldridge$^{2,3}$, J. C. Lee$^{4}$, S. Verley$^{1}$ \\
$^1$ Departamento de F\'\i sica Te\'orica y del Cosmos, Universidad de Granada, Campus Fuentenueva, Granada, Spain. \\
$^2$ Institute of Astronomy, University of Cambridge, Madingley Road, Cambridge CB3 0HA. \\
$^3$ Department of Physics, University of Auckland, Private Bag 92019, Auckland, New Zealand. \\
$^4$ Space Telescope Science Institute, 3700 San Martin Drive, Baltimore, MD 21218, USA.
} 

\begin{document}

\date{}

\pagerange{\pageref{firstpage}--\pageref{lastpage}} \pubyear{2002}

\maketitle

\label{firstpage}

\begin{abstract}
We present observational evidence that leakage of ionising photons from star-forming regions can affect the quantification of the star formation rate (SFR) in galaxies. This effect could partially explain the differences between the SFR estimates using the far ultraviolet (FUV) and the \ha\ emission. We find that leakage could decrease the SFR(\ha)/SFR(FUV) ratio by up to a  25 per cent. The evidence is based on the observation that the SFR(\ha)/SFR(FUV) ratio is lower for objects showing a shell \ha\ structure than for regions exhibiting a much more compact morphology. The study has been performed on three object samples: low luminosity dwarf galaxies from the Local Volume Legacy survey and star-forming regions in the Large Magellanic Cloud and the nearby Local Group galaxy 
M\,33. For the three samples we find differences ($1.1-1.4\sigma$) between the SFR(\ha)/SFR(FUV) for compact and shell objects. Although leakage cannot entirely explain the observed trend of SFR(\ha)/SFR(FUV) ratios for systems with low SFR, we show the mechanism can lead to different SFR estimates when using \ha\ and FUV luminosities. Therefore, further study is needed to constrain
the contribution of leakage to the low SFR(\ha)/SFR(FUV) ratios observed in dwarf galaxies and its impact on the \ha\ flux as a SFR indicator in such objects.
\end{abstract}


\section{Introduction}
Measuring the star formation rate (SFR) in galaxies has long been a major field of study in astronomy.  
Two of the most widely indicators of the star formation are the \ha\ recombination emission line and the 
UV non-ionising continuum. The \ha\ emission is produced by the recombination of the hydrogen ionised by the 
most massive (M\,$\ga$\,20\,$\msun$) and short-lived (t\,$\la$\,1-10\,Myr) stars. 
The UV emission primarily originates in the photospheres of lower mass (M\,$\ga$\,3\,$\msun$) and long-lived stars 
(t\,$\la$\,200\,Myr). Despite the different average timescales in the stellar lifetimes both 
indicators have been probed to trace well the SFR in galaxies over a range of 
luminosities and environments \citep{Kennicutt:1998p611}. 

The robustness of both indicators has been questioned for objects whose star formation history consists of a set of starburst episodes rather than a constant SFR \citep{Sullivan:2004p707,IglesiasParamo:2004p769,Boselli:2009p704,Bell:2001p721}. Recently, \citet{Lee:2009p609} showed that, in a complete sample of $\sim$300 star-forming galaxies within 11~Mpc volume, extinction corrected \ha\ and  far ultraviolet (FUV) emission give consistent SFRs for normal spiral galaxies (SFR $\approx 1\msun$ yr$\me$). However, for low luminosity dwarf galaxies (SFR $\approx 0.1\msun$ yr$\me$) \ha\ emission tends to under predict the SFR relative to the prediction of FUV emission. For galaxies with SFR $\la 0.003\msun$ yr$\me$ the ratio of SFR(\ha)/SFR(FUV) is on average lower than expected by a factor of two. 

Several explanations have been suggested for explaining this discrepancy \citep[see e.g.][for a review]{Meurer:2009p608,Lee:2009p609}: 
the amount of dust within star-forming regions, star formation history (SFH), porosity of the interstellar medium, stochasticity of the Initial Mass Function (IMF) and IMF variations. 

The differences in the SFR derived from \ha\ and FUV for low star-forming systems are present even when Galactic foreground  extinction corrections have been applied to the observed luminosities. For normal spiral galaxies \citep[see also][]{Botticella:2011p776} the SFR derived from FUV tend to be lower than the one 
derived using \ha\ emission, but after accounting for internal attenuation agreement is found using both tracers. However, for low star-forming systems internal attenuation increases the discrepancies between the two SFR estimates \citep{Meurer:2009p608,Lee:2009p609}. 

A departure in the assumption of constant star formation implied in the calibration relations can affect the value of SFR derived from FUV or \ha\ emission: for a galaxy with systematic star formation bursts the deficiency of ionising short-lived stars relative to the longer-lived lower mass stars emitting in the FUV will produce an \ha-to-FUV ratio lower than expected from a constant SFH \citep{Sullivan:2004p707,IglesiasParamo:2004p769}. However, to explain the observed variations in the SFR ratio the low star-forming systems must have very intense (a factor of 100) and very long ($\sim$100\,Myr) bursts of star formation \citep{IglesiasParamo:2004p769}. Such bursts of star formation do not seem to occur in these systems based on SFHs reconstructed from resolved stellar populations \citep{Weisz:2008p723,Mcquinn:2009p709}. 

Stochastic effects of an invariant IMF \citep[e.g][]{Cervino:2004p770,Corbelli:2009p781} or variation of the maximum stellar mass that can be formed in a stellar cluster leading to an integrated galactic initial mass function (IGIMF) \citep{Weidner:2005p800,PflammAltenburg:2009p675}, have been proposed to explain the differences in the observed SFR(\ha)/SFR(FUV) ratios. 
The assumption of an IGIMF describing the stellar population of the galaxy seems to produce results that explain the observations reasonably well, but it underestimates the \ha\ luminosity in systems with low SFR activity \citep{Fumagalli:2011p771,Weisz:2012p784}. Besides, the variation of an upper mass limit of the IMF is still under discussion, as there is some evidence showing that this variation does not exist \citep{Calzetti:2010p745}.  \citet{Eldridge:2011p755} and \citep{Fumagalli:2011p771} showed that stochastic IMF sampling combined with a cluster mass function and a cluster age distribution can well explain the differences in the SFR(\ha)/SFR(FUV) ratios. Using models of different SFHs and a fully populated IMF, \citet{Weisz:2012p784} compared the predictions of SFR(\ha) and SFR(FUV) with the observations and were also able to reproduce the decline of the observed SFR(\ha)/SFR(FUV) for lower luminosity systems. However, using the same models,  \citet{Weisz:2012p784} were not able to find a strong agreement between the \ha\ equivalent width or the UV luminosity modelled distributions and the observed ones.  

One of the other plausible explanations for the observed SFR ratio is the leakage of Lyman continuum photons from star-forming regions. In the conversion of \ha\ luminosity to star formation, it is assumed that all ionising radiation is absorbed by the neutral hydrogen in the star-forming regions and therefore every Lyman continuum photon will produce an \ha\ photon. However, it is known that some of the ionising photons can escape the regions of star formation \citep[e.g.][]{Oey:1997p612, Relano:2002p97,Eldridge:2011p756} and ionise the diffuse interstellar gas (DIG) in normal galaxies \citep{Zurita:2000p748,Zurita:2002p747,Wood:2010p801}. \citet{Oey:2007p749} have shown that the fraction of diffuse ionised gas seems to be higher in dwarf galaxies than that observed in normal galaxies. However, whether this translates into a difference in the escape fraction of ionising photons for each type of galaxy remains uncertain \citep{Hanish:2010p750}. If all the ionising photons escaping from the star-forming regions are able to ionise the DIG they should be included when integrating the total \ha\ emission from the whole galaxy. Therefore, for a constant SFH, we should not expect discrepancies between the SFR estimates from the total \ha\ or FUV luminosity of galaxies. However, if there are ionising photons 
escaping the galaxy disk or those ionising the DIG are not detected \citep{Melena:2009p775,Hunter:2010p713}, then we expect to find discrepancies between the SFRs obtained from \ha\ and FUV luminosities. Recently, \citet{Eldridge:2011p755} has shown that leakage of ionising photons in combination with pure stochastic IMF sampling better reproduce the observed SFR(\ha)/SFR(FUV) ratio at very low luminosities. 

There is some evidence that the \hii\ regions with \ha\ shell morphology (normally called bubbles and superbubbles) have in general a higher escape fraction of ionising photons \citep{Oey:1997p612}, based on the idea that once the shells are formed the ISM can fragment and produces voids of gas where the ionising radiation can escape \citep{Dove:2000p753}. \citet{Hoopes:2000p594} tried to apply this idea to a sample of \hii\ regions in M\,33 and found no correlation of compact or diffuse star-forming regions with the expected leakage fraction of ionising photons. Recently, \citet{Grossi:2010p754} found that the star-forming regions in M\,33 associated with younger and lower-mass stellar clusters seem to have higher escape fractions of ionising photons. 

In view of the renewed interest in the consistency of \ha\ and UV-based SFRs we have undertaken a more comprehensive 
examination of the possible role of ionising photon leakage from \hii\ regions. We present here observational evidence that even if the fraction of ionising photons escaping from galaxies and star-forming regions is not able to completely explain the observational trends by itself, it should be taken into account when trying to explain the observed phenomenology in combination with other mechanisms. 
We estimate here that leakage could decrease significantly the SFR(\ha)/SFR(FUV) ratio. Therefore, in order to explain the differences in the quantification of the SFR based on FUV and \ha\ luminosities, one should constrain the magnitude of leakage from star-forming regions.

Our study is based on a morphological classification of the \hii\ regions in low surface brightness dwarf galaxies in the Local Volume Legacy (LVL) survey sample \citep{Dale:2009p765,Lee:2011p766}. 
Using the \ha\ images from the LVL archive, we perform the morphological classification of the galaxies based on the morphology of the most luminous \hii\ regions in each galaxy.  The same morphological discrimination is applied on \ha\ images of individual regions in the Large Magellanic Cloud (LMC) and the nearby galaxy M\,33. The most luminous \hii\ regions show \ha\ luminosities and SFRs comparable to the low surface brightness dwarf galaxies. Besides, the star formation for these galaxies occur in the form of short-lived burst episodes similar to those in the \hii\ regions \citep[see e.g.][]{Gerola:1980p780,Weisz:2012p784}, therefore it is reasonable to study and compare both object samples. In Section\,2 we explain the data for each sample and the morphological classification we have applied to the objects. In Section\,3 we estimate the SFR(\ha)/SFR(FUV) for each object in the samples and show the trends with the \ha\ morphology. We analyse the results and show the importance of the leakage effect in the observational trends in Section\,4 and the conclusions are presented in Section\,5. 

\section{Methodology}
\subsection{Local Volume Legacy galaxy sample}\label{LVLsec}
\begin{figure*}
\includegraphics[width=\textwidth]{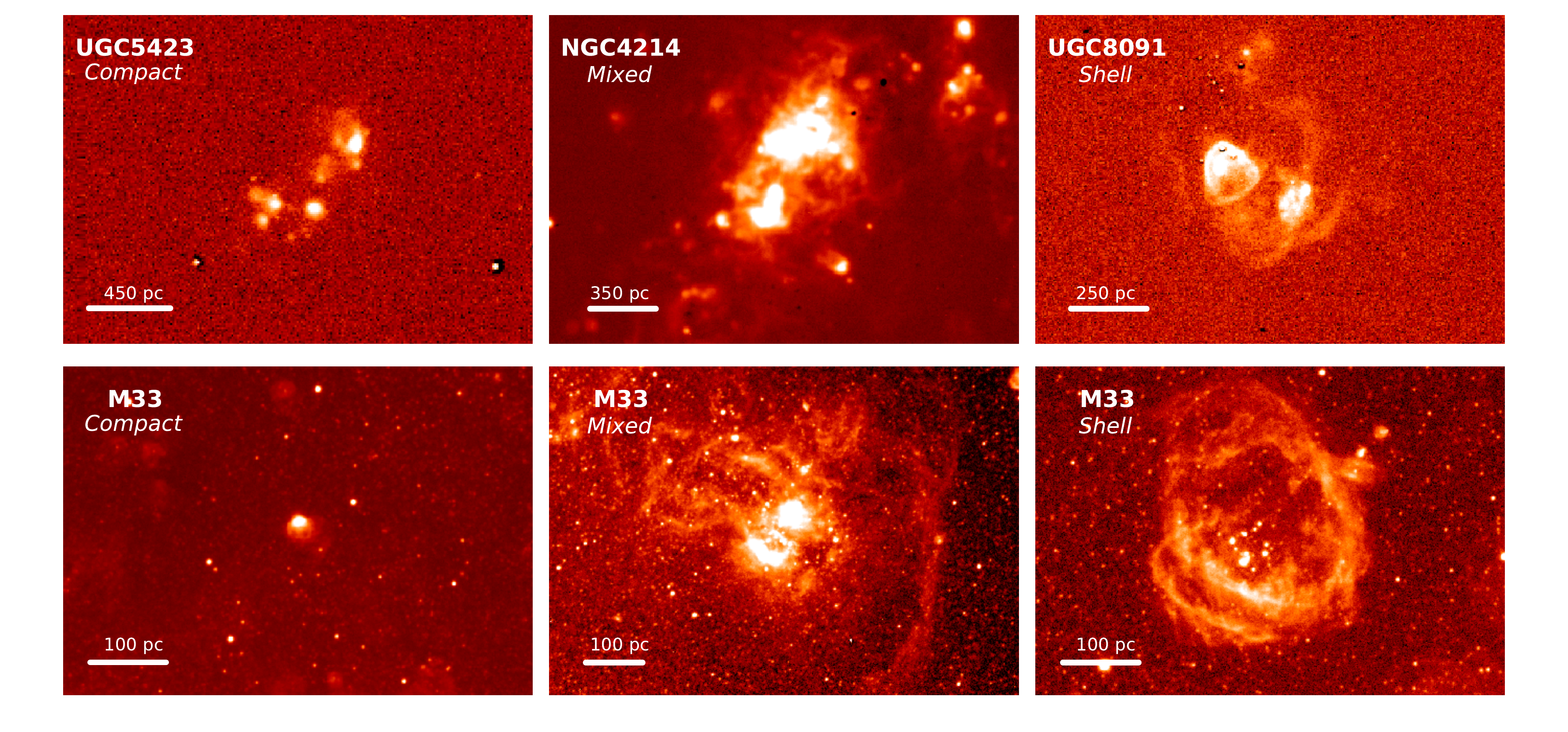}
\caption{ \ha\ images for examples of compact, mixed and shell objects: galaxies in the LVL sample are shown in the top row,  while the star-forming regions in M\,33 are shown in the bottom row.}
\label{example}
\end{figure*}

\begin{figure*}
\includegraphics[width=\textwidth]{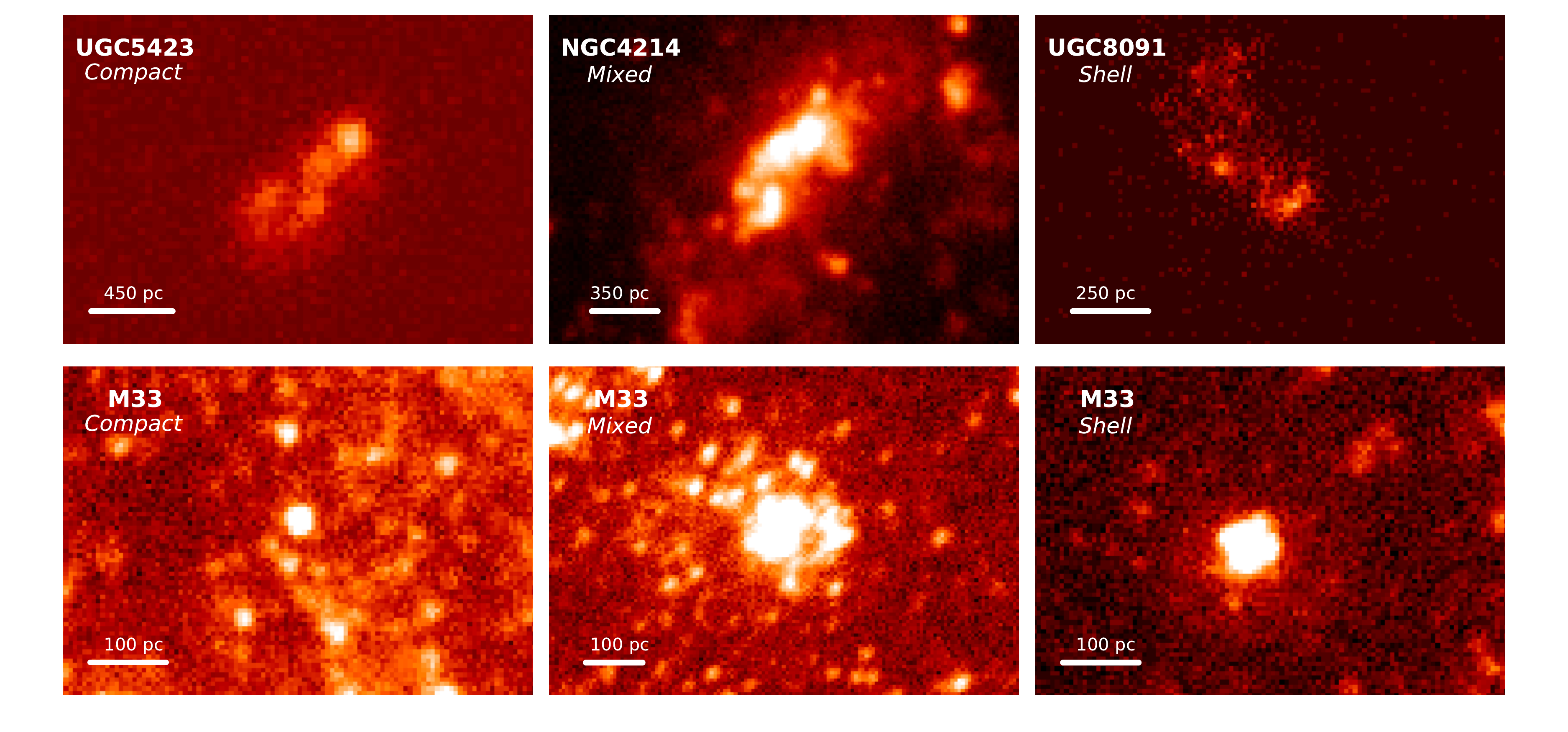}
\caption{FUV images of the same objects as those shown in Fig.~\ref{example}: galaxies in the LVL sample are shown at the top, while the star-forming regions in M\,33 are shown in the bottom row.}
\label{example_fuv}
\end{figure*}

We classified the LVL galaxies from the Data Release 5\footnote{$http://irsa.ipac.caltech.edu/data/SPITZER/LVL/$}  \citep{Lee:2009p609} in terms of morphology using the \ha\ images available in the archive. We chose only galaxies with RC3 type T 10 and 11 to avoid grand design spirals. The classification was done visually by one of the authors (MR) in terms of morphology of the \hii\ regions: {\it compact} are galaxies presenting compact knots (one or several), {\it mixed} 
are galaxies presenting compact knots and filamentary structures, and {\it shells} are galaxies with 
clear shells (one or several). \ha\ images of some examples of each classification are shown in the top row of Fig.~\ref{example}, while the FUV images of the same objects are shown in the top row of Fig.~\ref{example_fuv}. 
The integrated \ha\ and FUV fluxes for the LVL galaxy sample were presented in \citet{Kennicutt:2008p678} and in \citet{Lee:2010p708}, respectively. 

The comparison of the \ha\ and FUV SFRs for the galaxy sample was done in \citet{Lee:2009p609} showing that for low luminosity dwarf galaxies \ha\ luminosity tends to underpredict the SFR relative to the FUV luminosity. The trend is seen not only for Galactic extinction corrected SFRs but also when internal extinctions were applied. The \ha\ extinction was derived using the Balmer decrement while the extinction in the FUV was obtained using the total infrared (TIR) to FUV flux ratio and applying models from \citet{Buat:2005p667}. For those objects with no available TIR emission estimates scaling relations were used to derive the FUV extinction. Besides, scaling relations were used when Balmer decrements or 
TIR/FUV ratios were not available \citep[see][]{Lee:2009p609}.

\subsection{Sample of star-forming regions in M\,33}\label{M33sec}
We selected a sample of star-forming regions in the nearby galaxy M\,33. This galaxy is a particularly  
suitable object for this study because its distance \citep[840~kpc;][]{Freedman:1991p485} 
allowed us to observe the \hii\ regions as individual objects and to perform a morphological classification of the sample.
Besides, the large number of star-forming regions in the disk of M\,33 allowed us to create a significant statistical sample of 
objects. Some clear shells have been observed in several wavelength ranges from the optical to the Far Infrared (FIR) with the recent 
observations from Herschel \citep{Kramer:2010p688,Verley:2010p687}. An example is shown in the lower-right corner of 
Fig.~\ref{example}, this shell is also clearly revealed in the FIR bands from Herschel observations \citep{Verley:2010p687} and will be studied in detail in \citet[][in prep.]{Relano:2012}. 

We used the continuum-subtracted \ha\ image of M\,33 from \citet{Hoopes:2000p594} to select our sample. The angular resolution of this image, 5~\arcsec, corresponding to a linear scale of $\sim$20\,pc, is good enough to identify the \hii\ region sample. The \hii\ region selection was done visually by one of the authors (MR), as the aim was to find a set of relatively isolated 
objects with a clear morphology which facilitates the classification \citep[see also][in prep.]{Relano:2012}.
The morphological classification was done using the wide-field KPNO mosaic \ha\ images 
from the Local Group Galaxy Survey (LGGS) collaboration \citep{Massey:2006p517}. These \ha\ images provide a much 
better spatial resolution ($\sim$0.8\,arcsec, corresponding to $\sim$4\,pc) allowing a more accurate morphological classification. Our final sample consists of 117 objects. We applied the same criteria for the morphological classification as described in section\,\ref{LVLsec} and found 10 compact 
regions, 46 mixed regions and 61 regions with shell morphology.
 
We performed photometry in the FUV, \ha, 24\,\mi\ and 8\,\mi\ bands using individual apertures for each \hii\ region. The FUV image was taken from the GALEX \citep{Martin:2005p691,GildePaz:2007p692} data archive. The angular resolution for this image was $\sim$4\arcsec. 
The 8\,\mi\ and 24\,\mi\ emissions were obtained from the IRAC and MIPS ({\it Spitzer}) images with $\sim$3\arcsec\ and 6\arcsec\ resolution, respectively. The data reduction of the {\it Spitzer} images is explained in \citet{Verley:2007p574}. As the images were used to obtain the TIR luminosity of the objects, we needed to eliminate the stellar contribution in these bands. We used the IRAC-3.6\,\mi\ image and followed the relation proposed by Helou et al. (2004) for this purpose. Finally, all the images were smoothed to a common 6\arcsec\ resolution (the 24\,\mi\ band resolution, which is the lowest one for the images of M\,33 used in this study) and registered to the same pixel size as the \ha\ image from \citet{Hoopes:2000p594}. 

\begin{figure*}
\includegraphics[width=0.45\textwidth]{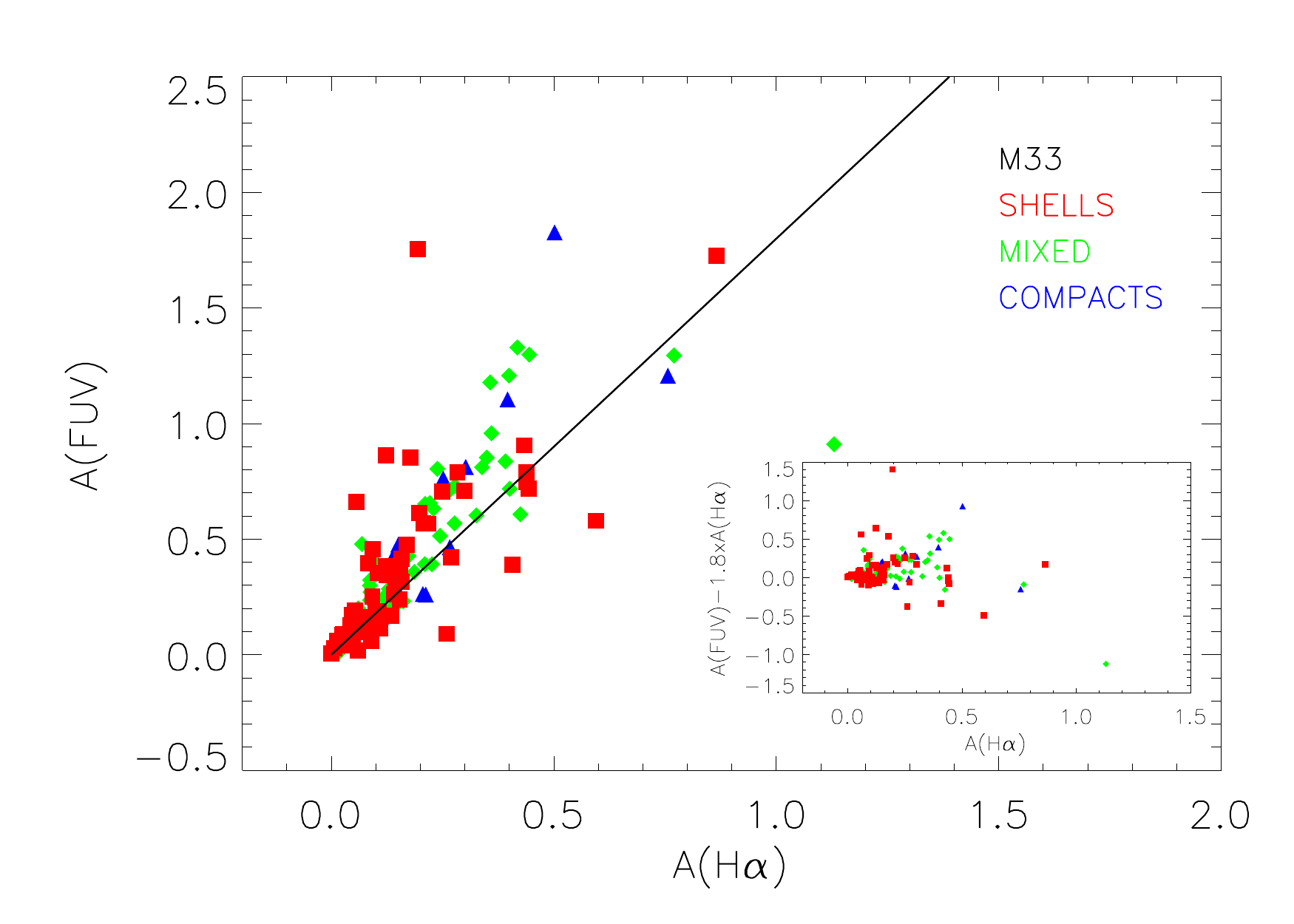}
\includegraphics[width=0.45\textwidth]{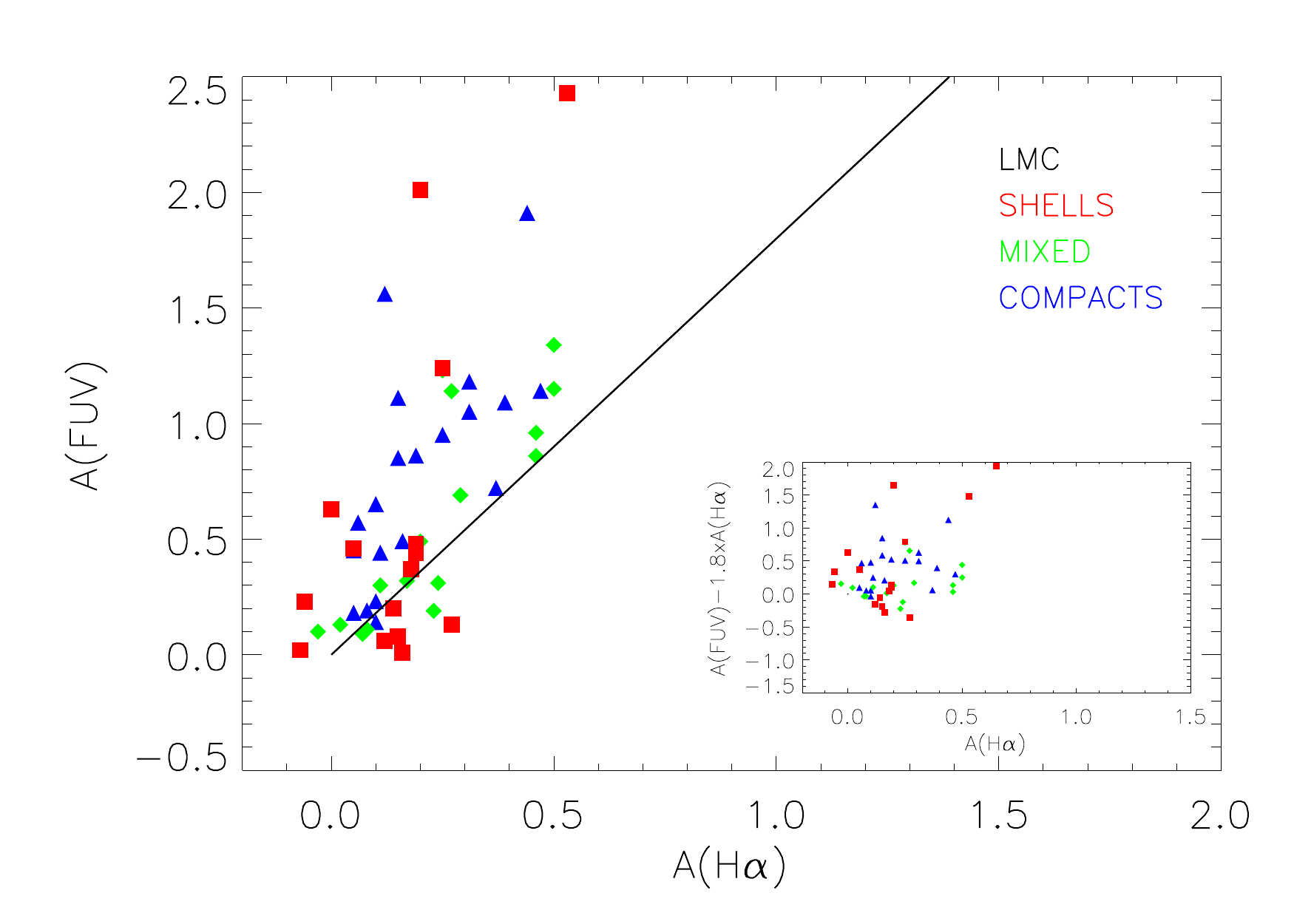}
\caption{Left: FUV extinction versus \ha\ extinction for the \hii\ region sample in M\,33. The continuous line represents the relation A(FUV)$=1.8\times$A(\ha) derived using the Calzetti's reddening curve. The \ha\ extinction was obtained using the \ha\ /24\mi\ ratio. Right: Same figure but for the \hii\ region sample in LMC. The \ha\ extinction was obtained using the \ha\ /\hb\ emission line ratio. In the lower-right corner of each plot we show the difference between A(FUV) and the values expected following Calzetti's reddening curve.}
\label{ext}
\end{figure*}

The \ha\ photometry was performed on the Hoopes \& Walterbos image, as the \ha\ image of the LGGS collaboration presented some saturated zones in the brightest part of the most luminous \hii\ regions. The FUV and \ha\ fluxes were corrected for Galactic extinction using $E(B-V)=0.07$ \citep{vandenBergh:2000p502} and \citet{Cardelli:1989p595} extinction law. The internal \ha\ extinction for each region was obtained using the \ha\ and 24\,\mi\ fluxes and 
assuming that the absorbed \ha\ luminosity in the region scales with the 24\,\mi\ luminosity with a factor of $a=0.031$ 
\citep{Kennicutt:2007p488}. FUV extinction was obtained using the empirical relation between the ratio of the TIR 
to FUV luminosities and the UV spectral index $\beta$ following the formalism given in 
\citet{Calzetti:2001p661}. The TIR emission was obtained from the linear combination of 
8\,\mi\ and 24\,\mi\ fluxes given in Table~1 in \citet{Boquien:2010p674} corresponding to regions with 45\arcsec\ aperture size. 
We compare the \ha\ and FUV extinctions in the left panel of Fig.~\ref{ext}. For most of the \hii\ regions the 
relation between both extinctions agrees well with the expected relation (A(FUV)$=1.8\times$A(\rm \ha)) 
from Calzetti's reddening curve. The values of the FUV extinction are within the range obtained by \citet{Verley:2009p573}.

\subsection{\hii\ region sample in the Large Magellanic Cloud}
Our \hii\ region selection for the LMC was determined by the sample presented 
in \citet{Bell:2002p677}, which represents a subsample of isolated regions from \citet{Caplan:1986p663}. 
\citet{Bell:2002p677} presented the fluxes at \ha\ and 1500\,\AA\ for a set of \hii\ regions in the LMC within 
a diameter aperture of 4.9\arcmin, corresponding to 65\,pc at the distance of LMC \citep[45.9\,kpc;][]{Fitzpatrick:2002p741}. 
We performed the morphological classification using the LMC \ha\ image from \citet{Kennicutt:1986p772} with an angular resolution image of $\sim$7\,arcsec. The classification was done visually by one of the authors (MR).  We applied the same criteria as those for the LVL and the M\,33 sample and catalogued the \hii\ region sample. We found 20, 16 and 16 compact, mixed and shell regions, respectively.  

Foreground Galactic extinction correction was applied to both \ha\ and FUV (1500\,\AA) fluxes using the Galactic reddening 
E(B-V)=0.06 given in \citet{Bell:2002p677}. These authors also estimate the internal \ha\ and FUV extinction for each region. 
The former is based on the Balmer decrement and the latter on the TIR/FUV ratio with the calibrations derived in \citet{Gordon:2000p742}. We compare both extinctions in the right panel of Fig.~\ref{ext}. Most of the regions show higher values of FUV extinctions than those derived using Calzetti's extinction curve. However, as we study here the \hii\ regions from M33 and LMC in a separated way the fact that we apply different extinction laws for the two samples does not affect the final results of this paper.

\section{Results}

\begin{figure}
\begin{center}
\vspace{-2cm}
\hspace{-0.5cm}
\includegraphics[width=0.8\columnwidth,angle=90]{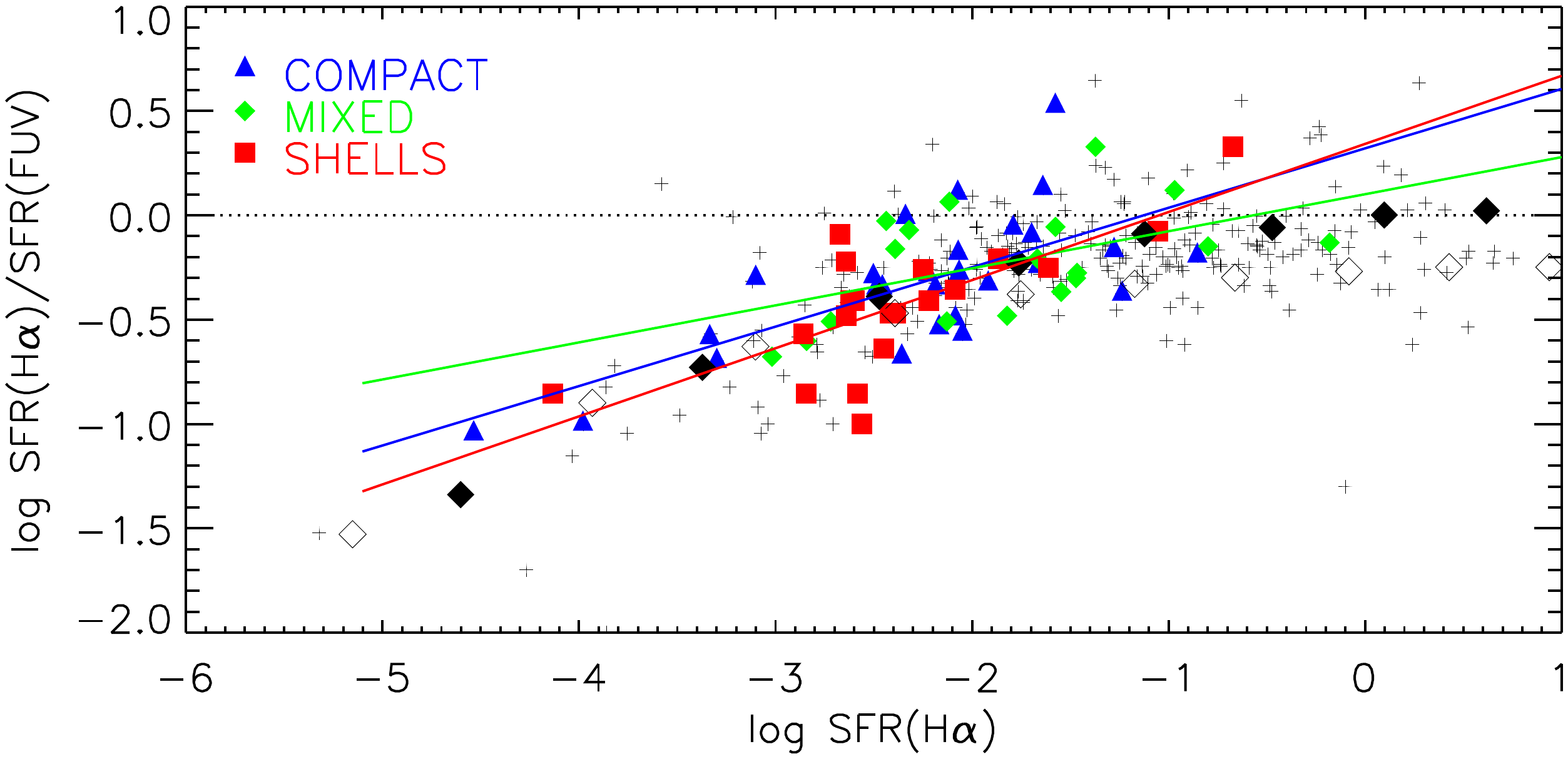}    
\caption{Logarithm of the SFR(\ha)/SFR(FUV) versus logarithm of the SFR(\ha) for the LVL galaxy sample (from Fig.~5 in \citealt{Lee:2009p609}) with colour-code in terms of morphology. Blue: compact galaxies, green: galaxies catalogued as mixed, and red: galaxies with \ha\ shell morphology. A least-square linear fit is shown for each morphology. The open and compact black diamonds correspond to the standard and minimal models of the IGIMF from \citet{PflammAltenburg:2009p675}.}
\label{leefig5}
\end{center}
\end{figure}

In Fig.~\ref{leefig5} we reproduce figure~5 from \citet{Lee:2009p609} using a colour-code for each classified galaxy:  {\it compact} are blue, {\it mixed} are green and red are {\it shells}. In general, the galaxies classified as mixed are slightly more luminous than the 
compact or shells: $\langle\log \rm{L}(\ha/\rm {erg\,s\me})\rangle=38.8\pm0.2$, $\langle\log \rm{L}(\ha/\rm {erg\,s\me})\rangle=39.3\pm0.2$, $\langle\log \rm{L}(\ha/\rm {erg\,s\me})\rangle=38.7\pm0.2$ for compact, mixed, and shells, respectively. This is expected as the mixed galaxies are formed by several intense knots of star formation intertwined with some diffuse emission. The mean value for the $\log$(SFR(\ha)/SFR(FUV)) ratio in galaxies classified as mixed is higher than for the compact and shells, the shells having the lowest ratio: $\langle\log$(SFR(\ha)/SFR(FUV))$\rangle$ is $-0.32\pm0.06$, $-0.22\pm0.06$ and $-0.43\pm0.07$ for compact, mixed and shells, respectively. In an attempt to quantify this trend we have performed a linear fit to the data in Fig.~\ref{leefig5} for each classification. The fits are shown in the figure with the corresponding colour line for each morphology: blue for compacts, green for mixed and red for shells. The results of the fit are presented in Table~1. The linear fit for the shell morphology has the steepest slope, while the mixed show the shallowest one. We will discuss these results in section~\ref{discuss}.

\begin{figure}
\begin{center}
\vspace{-2cm} 
\hspace{-0.5cm} 
\includegraphics[width=0.8\columnwidth,angle=90]{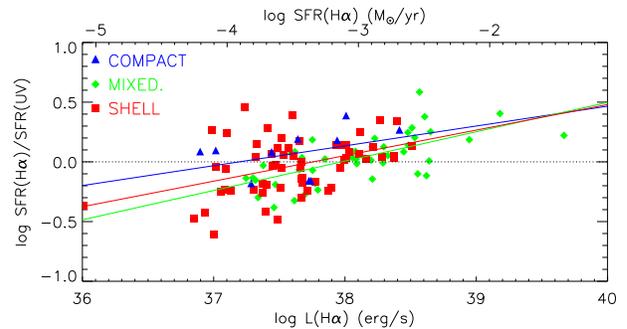}
\caption{Logarithm of SFR(\ha)/SFR(FUV) versus logarithm of the \ha\ luminosity for the \hii\ regions in M\,33 sample. Extinction corrections to \ha\ and to FUV luminosities have been applied.  A least-square linear fit is shown for each morphology classification.}
\label{m33fig}
\end{center}
\end{figure}

We used the \ha\ and FUV extinction derived in the previous section to correct the observed \ha\ and FUV fluxes 
of the \hii\ regions in M\,33. Then, we converted them into SFRs using the relations given in 
\citet{Kennicutt:1998p611}. These relations assume a constant SFRs over $\approx$100\,Myr and 
are suitable for galaxies rather than for \hii\ regions caused by short-lived bursts of star formation. However, we still 
apply these relations in order to compare the results with the LVL sample. Since we are using the ratio between the two SFR calibrators our conclusions will not change if we use other calibrations more suitable for \hii\ regions. In Fig.~\ref{m33fig} we present the logarithm of SFR(\ha)/SFR(FUV) as a function of the logarithmic \ha\ luminosity for the \hii\ regions in M\,33. The colour code is the same as in Fig.~\ref{leefig5}. The mixed regions are in general more luminous than the compact regions 
and the shells: mean values for the logarithmic \ha\ luminosity for each class are: $\langle\log \rm{L}(\ha/\rm {erg\,s\me})\rangle=37.6\pm0.1$, $\langle\log \rm{L}(\ha/\rm {erg\,s\me})\rangle=38.1\pm0.2$ and $\langle\log \rm{L}(\ha/\rm {erg\,s\me})\rangle=37.6\pm0.1$ 
for compact, mixed and shells, respectively. 
The mean values of the SFR ratios, $\langle\log$(SFR(\ha)/SFR(FUV))$\rangle$, are $0.07\pm0.06$, $0.03\pm0.03$ and, $-0.03\pm0.03$ for compact, mixed, and shells, respectively. The difference between $\langle\log$(SFR(\ha)/SFR(FUV))$\rangle$ for compact and shells is $0.10\pm0.07$, a marginal difference of   
$1.4\sigma$. The continuous lines in Fig.~\ref{m33fig} show the linear fit to the data in each classification (see Table~1 for fitting coefficients).


\begin{figure}
\begin{center}
\vspace{-2cm} 
\hspace{-0.5cm}
\includegraphics[width=0.8\columnwidth,angle=90]{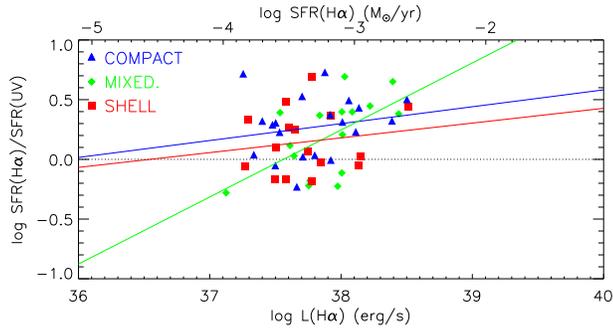}     
\caption{Logarithm of SFR(\ha)/SFR(FUV) versus logarithm of the \ha\ luminosity for the \hii\ regions in LMC sample. Extinction corrections derived from \citet{Bell:2002p677} have been applied. A least-square linear fit is shown for each sample.} 
\label{lmcfig}
\end{center}
\end{figure}

\ha\ and FUV fluxes for the regions in the LMC have been corrected from the extinction derived in \citet{Bell:2002p677} and SFR calibrations have been applied to derived 
the corresponding SFRs using the relations in \citet{Kennicutt:1998p611}. The results are shown in Fig.~\ref{lmcfig}, as well as the linear fits for each classification (see Table~1 for fitting coefficients). For this sample, there is also a marginal separation between the $\log$(SFR(\ha)/SFR(FUV)) ratio for compact and for shell regions: $\langle\log$(SFR(\ha)/SFR(FUV))$\rangle$ are $0.27\pm0.06$, $0.19\pm0.08$ and $0.15\pm0.07$ for compact, mixed and shells, respectively. The difference in the mean values between compacts and shells is 0.12$\pm$0.09, a $1.3\sigma$ difference.

We have not combined the \hii\ regions of M 33 and LMC because the 
samples were not created using the same procedure. For M 33 we used the \ha\ image from \citet{Hoopes:2000p594} 
and identified those isolated \hii\ regions with clear morphology. For LMC, we did not create the sample, we rather 
performed the morphological classification of a set of \hii\ regions previously identified with the purpose of studying the Balmer extinction in each 
region \citep[see][]{Caplan:1986p663}. Therefore, the \hii\ regions in LMC are biased to the most luminous ones 
emitting at \ha\ and \hb. This bias can produce systematic differences in the \ha\ luminosity ranges 
between M 33 and LMC samples that can affect the comparison when the combined sample is studied as a single set.

\begin{table}
\begin{center}
\caption{Linear fits for the data: $\log (\rm{SFR}(\ha)/ \rm{SFR(FUV)})$$=a$$\times\log (\rm{L}(\ha))+$$b$.}
\begin{tabular}{ccccc}
\hline\noalign{\smallskip}
Sample &  Type & $a$ & $b$ &  correl. coeff. \\  
\hline
\hline

LVL & compact & $0.28\pm0.06$ & $0.3\pm0.1$ & $0.70$  \\
LVL & mixed & $0.18\pm0.08$ & $0.1\pm0.2$ & $0.50$  \\
LVL & shells & $0.33\pm0.07$ & $0.3\pm0.2$ &  $0.73$  \\
\hline
M\,33 & compact & $0.2\pm0.1$ & $-6\pm5$ & $0.40$ \\
M\,33 & mixed & $0.25\pm0.04$ & $-9\pm2$  & $0.67$ \\
M\,33 & shells & $0.21\pm0.06$ & $-8\pm2$ & $0.43$ \\
\hline
LMC & compact & $0.1\pm0.2$ & $-5\pm6$ & $0.19$ \\
LMC & mixed & $0.6\pm0.2$ & $-21\pm8$ & $0.58$  \\
LMC & shells & $0.1\pm0.2$ & $-5\pm8$ & $0.15$  \\
\hline
\hline\noalign{\smallskip}
\end{tabular}
\label{lambdaint}
\end{center}
\end{table}

\begin{figure}
\begin{center}
\includegraphics[width=\columnwidth]{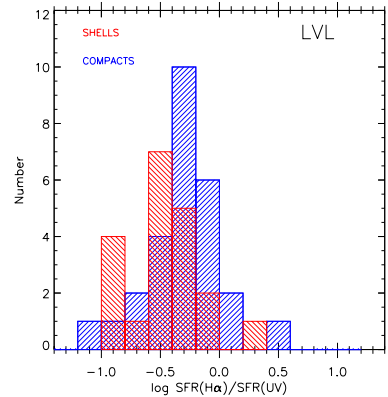} 
\vspace{0.5cm}
\includegraphics[width=\columnwidth]{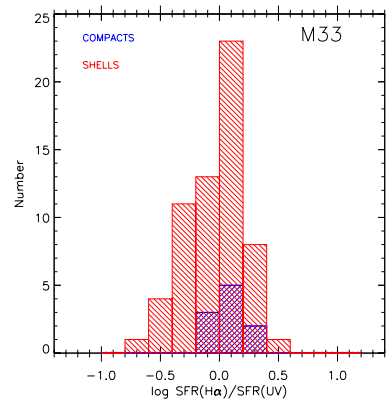} 
\vspace{-1.5cm}
\vspace{0.2cm}
\includegraphics[width=\columnwidth]{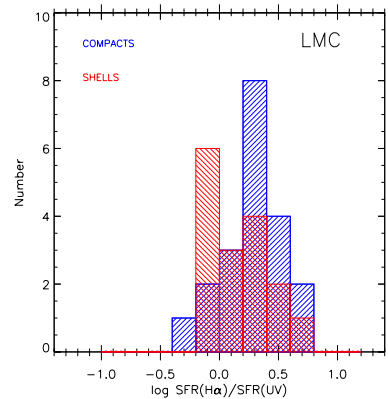} 
\vspace{-1cm}
\caption{Histograms of the $\log$(SFR(\ha)/SFR(FUV)) for the three samples. All data are extinction corrected. Red histograms are for shells and blue for compact regions.}
\label{histo}
\end{center}
\end{figure}

We quantify the distribution of the $\log$(SFR(\ha)/SFR(FUV)) ratios using histograms for the LVL, M 33 and LMC samples (Fig.~\ref{histo}, upper, mid, and bottom panels, respectively). We only show the compact and shell classifications to clarify the results. The bins in $\log$(SFR(\ha)/SFR(FUV)) are 0.2~dex for the three sets. 
The separation between compacts and shells for the LVL, M 33, and LMC samples are $0.11\pm0.10$, $0.10\pm0.07$, and $0.12\pm0.09$. In the three samples we find marginal separations, $\sim1.1-1.4\sigma$, for the compact and shell distributions.

\section{Discussion}\label{discuss}

In the previous section we have presented the trends of the SFR(\ha)/SFR(FUV) for each data sample: the dwarf galaxies with moderate SFR from the LVL sample and the set of \hii\ regions in the nearby M\,33 and LMC. Most of the objects in the three samples are in the range of low SFR regime, $-5~<~\log\rm{SFR}(\ha)/(\msun\rm{yr\me})<-1$. In each sample the mean values of $\log$(SFR(\ha)/SFR(FUV)) are lower for the objects catalogued as shells than for the rest of the objects, either mixed or compact regions. For the LVL sample we find a difference in the $\log$(SFR(\ha)/SFR(FUV)) for compact and shells of $0.11\pm0.10$ and for M\,33 and LMC samples we find differences of $0.10\pm0.07$ and $0.12\pm0.09$.

If the shell morphology of star-forming regions favours the existence of zones of low density gas via holes or fragmentation of the shell walls, and the ionising photons are indeed more likely to leak the star-forming regions between the fragments, then the difference in $\log$(SFR(\ha)/SFR(FUV)) between compact and shells could be due to the leakage of ionising photons. 

It is interesting to note that \citet{Hoopes:2000p594} did not find a correlation between the diffuse and ring \hii\ regions in M 33 and the fraction of ionising photons escaping the regions. However, these authors are restricted to regions in M 33 for which they were able to study the stellar content. These \hii\ regions have low luminosity (only a few are more luminous than 10$^{38}$\ergs after extinction correction has been applied) and they represent a small fraction of the \hii\ region population in M 33 \citep[see fig.\,1 of][]{Hoopes:2000p594}. In order to infer firm conclusions 
about whether leakage is more important in shells and ring-like regions than in compact objects, a detailed study of the stellar content of \hii\ regions with different morphology would be needed.

Assuming that compact regions are able to absorb all the ionising radiation 
--which would be a strong case, as a compact region with a low filling factor 
would be able to leak radiation through the zones between clumps 
\citep[see][]{Giammanco:2004p626,Giammanco:2005p625}-- and 
the shells leak all the ionising radiation coming from the stellar cluster; 
then we could say that the difference between the $\log$(SFR(\ha)/SFR(FUV)) 
of compact and shells would be entirely due to a leakage effect. The maximum 
observed difference of 0.12 between the compact and shell regions translates into a factor of 0.76 between the 
SFR(\ha)/SFR(FUV) ratio. 
Therefore, we can say that leakage could change the SFR(\ha)/SFR(FUV) ratio 
between 0 to $\sim$25 per cent.

We can compare our results with those expected from model stellar 
populations. We use the single-star stellar populations of \citet{Eldridge:2009p778} 
that assume the \hii\ region stellar population is described by an 
instantaneous burst with a Chabrier IMF \citep{Chabrier:2003p783}. These predict a $\log$(L(\ha)/L(FUV)) of the 
order of 0.4 for an \hii\ region 4\,Myr old. However, for our observed \hii\ 
regions, $\log$(L(\ha)/L(FUV))$\approx$0.01-0.02, much lower than the model ratios. 
Therefore, leakage of ionising photons is expected if the age of the \hii\ 
regions are $\approx$\,4\,Myr. Indeed, for NGC 604 and for the set of \hii\ 
regions from \citet{Oey:1997p612} included in our LMC sample we observe 
$\log$(L(\ha)/L(FUV))$\sim$0.01-0.02, consistent with a leakage fraction
($\sim$0-50 per cent) estimated for these regions 
\citep[][for NGC~604, and \citealt{Oey:1997p612} for the LMC \hii\ regions]{Eldridge:2011p756}.

The models only predict ratios of the order of 0.01 at an age of 10\,Myr. It 
is unlikely for compact \hii\ regions emitting at \ha\ to be this old, but 
shell-like \hii\ regions are expected to have ages greater than 4\,Myr and the assumption of no 
leakage could be plausible. Recently, \citet{Whitmore:2011p726} have shown evidence of the 
relation between the region morphology and the age of the central stellar cluster: 
very young (a few\,Myr) clusters show the \ha\ emission of the ionised gas coincident with the cluster stars, for slightly older clusters ($\approx$\,5\,Myr) the gas emission is located in small shell structures around the stars, and in still older clusters ($\approx$\,5-10\,Myr) the \ha\ emission shows even larger shell structures. They also show that if no \ha\ emission is associated with the cluster then this is generally older than $\approx$\,10\,Myr.
Following this study our sample of objects would be younger than 10\,Myr,  as they still 
exhibit \ha\ emission associated with the stellar clusters. Moreover, assuming a standard 
shell expansion velocity of 60\,km\,s$\me$ the 
radii of the shells at $t=10$\,Myr would be $\approx$600\,pc \citep{Relano:2005p644,Verley:2010p687}. Most of the observed 
\hii\ regions in our sample are smaller, therefore their age must also be less than 
10\,Myrs. 

We have run models to describe the behaviour of the $\log$(SFR(\ha)/SFR(FUV)) ratio 
as a function of age. For a 10$^5$\,M$_\odot$ stellar 
cluster, typical of the most luminous \hii\ regions, the $\log$(SFR(\ha)/SFR(FUV)) ratio  
decreases by a factor of more than one order of magnitude between the young 
(2-3\,Myr) and the more evolved (5-8\,Myr) \hii\ regions. For a coeval stellar population 
we would then expect a separation between compact and shells of one order of magnitude 
if age is the only responsible to explain the differences in $\log$(SFR(\ha)/SFR(FUV)). 

Other mechanism that can decrease the number of ionising photons is the absorption of Lyman 
continuum photons by the dust inside the regions \citep{McKee:1997p802,Hirashita:2003p768,IglesiasParamo:2004p769}. 
This mechanism will decrease the SFR(\ha)/SFR(FUV) ratio, as less ionising photons will be 
able to ionise the hydrogen. We would then expect that  
regions with lower SFR(\ha)/SFR(FUV) would have higher dust fractions; namely,  
shell regions would have higher fractions of dust than compact regions. This seems not plausible 
as shells normally have swept up the 
gas and dust and do not show in general emission at 24\,\mi\  \citep[see][]{Verley:2010p687}.

Despite of the intrinsic difficulties in quantifying the escape fraction of ionising photons from star forming regions, we show here that this mechanism has an effect in the observed trends for SFR(\ha)/SFR(FUV). Leakage mechanism, in combination with possible variations of the IMF: 
via stochasticity \citep{Cervino:2004p770}, random sampled IMF \citep{Corbelli:2009p781} or assuming an IGIMF \citep{PflammAltenburg:2009p675}, might well explain the differences in the SFR(\ha)/SFR(FUV) for star-forming objects. 

\section{Summary and conclusions}
We have studied the SFR for a sample of \hii\ regions in M\,33, LMC, and a set of dwarf galaxies in the LVL sample. We have classified them in compact, mixed, and shell regions and analysed the SFR(\ha)/SFR(FUV) ratio for the three morphology types. For M\,33 and LMC samples we obtained differences of $0.10\pm0.07$ and $0.12\pm0.09$ in the 
$\langle\log$(SFR(\ha)/SFR(FUV))$\rangle$ between compact and shell regions, a separation of $\sim1.4-1.3\sigma$. For the LVL sample we found a smaller difference, $\langle\log$(SFR(\ha)/SFR(FUV))$\rangle=-0.32\pm0.06$ for compacts and $\langle\log$(SFR(\ha)/SFR(FUV))$\rangle=-0.43\pm0.07$ for shells, corresponding to a difference of $0.11\pm0.10$.
The maximum 
observed difference of 0.12 between the compact and shell regions translates into a factor of 0.76 between the 
SFR(\ha)/SFR(FUV) ratio. 

We present here observational evidence that the escape of ionising photons from individual star-forming regions in M\,33, LMC, and a sample of dwarf galaxies might account for up to a 25 per cent decrease in the SFR(\ha)/SFR(FUV). Rather than performing a detailed quantification of the leakage fraction of ionising photons we show here that this mechanism should be considered when trying to explain the lower than expected SFR(\ha)/SFR(FUV) for low star-forming objects. It will probably be a combination of different mechanisms: leakage, variations of IMF, SFH, and age effects, which might finally explain the decrease of the SFR(\ha)/SFR(FUV) ratio for low luminosity systems.

\section{Acknowledgments}
We would like to thank C. Hao, B. Johnson and J. E. Beckman for their useful discussions. 
MR has been supported by a Marie Curie Intra European Fellowship within 
the 7th European Community Framework Programme and part of this research 
has been supported by the ERG HER-SFR from the EC.  This work was partially supported by a Junta de Andaluc\'ia Grant FQM108, a Spanish MEC Grant AYA-2007-67625-C02-02, and Juan de la Cierva fellowship Program.

\bibliographystyle{mn2e}
\bibliography{mnras} 

\begin{thebibliography}{}

\bibitem[\protect\citeauthoryear{Bell, Gordon, Kennicutt et~al.,}{Bell
  et~al.}{2002}]{Bell:2002p677}
Bell E.~F.,  Gordon K.~D.,  Kennicutt R.~C.,    et~al., 2002, ApJ, 565, 994

\bibitem[\protect\citeauthoryear{Bell \& Kennicutt}{Bell \&
  Kennicutt}{2001}]{Bell:2001p721}
Bell E.~F.,  Kennicutt R.~C.,  2001, ApJ, 548, 681

\bibitem[\protect\citeauthoryear{Boquien, Bendo, Calzetti et~al.,}{Boquien
  et~al.}{2010}]{Boquien:2010p674}
Boquien M.,  Bendo G.,  Calzetti D.,    et~al., 2010, ApJ, 713, 626

\bibitem[\protect\citeauthoryear{Boselli, Boissier, Cortese et~al.,}{Boselli
  et~al.}{2009}]{Boselli:2009p704}
Boselli A.,  Boissier S.,  Cortese L.,    et~al., 2009, ApJ, 706, 1527

\bibitem[\protect\citeauthoryear{Botticella, Smartt, Jr. et~al.,}{Botticella
  et~al.}{2011}]{Botticella:2011p776}
Botticella M.~T.,  Smartt S.~J.,  Jr. R. C.~K.,    et~al., 2011, arXiv,
  astro-ph.CO

\bibitem[\protect\citeauthoryear{Buat, Iglesias-P{\'a}ramo, Seibert
  et~al.,}{Buat et~al.}{2005}]{Buat:2005p667}
Buat V.,  Iglesias-P{\'a}ramo J.,  Seibert M.,    et~al., 2005, ApJ, 619, L51

\bibitem[\protect\citeauthoryear{Calzetti}{Calzetti}{2001}]{Calzetti:2001p661}
Calzetti D.,  2001, PASP, 113, 1449

\bibitem[\protect\citeauthoryear{Calzetti, Chandar, Lee et~al.,}{Calzetti
  et~al.}{2010}]{Calzetti:2010p745}
Calzetti D.,  Chandar R.,  Lee J.~C.,    et~al., 2010, ApJ, 719, L158

\bibitem[\protect\citeauthoryear{Caplan \& Deharveng}{Caplan \&
  Deharveng}{1986}]{Caplan:1986p663}
Caplan J.,  Deharveng L.,  1986, Astronomy and Astrophysics (ISSN 0004-6361),
  155, 297

\bibitem[\protect\citeauthoryear{Cardelli, Clayton \& Mathis}{Cardelli
  et~al.}{1989}]{Cardelli:1989p595}
Cardelli J.~A.,  Clayton G.~C.,    Mathis J.~S.,  1989, ApJ, 345, 245

\bibitem[\protect\citeauthoryear{Cervi{\~n}o \& Luridiana}{Cervi{\~n}o \&
  Luridiana}{2004}]{Cervino:2004p770}
Cervi{\~n}o M.,  Luridiana V.,  2004, A\&A, 413, 145

\bibitem[\protect\citeauthoryear{Chabrier}{Chabrier}{2003}]{Chabrier:2003p783}
Chabrier G.,  2003, PASP, 115, 763

\bibitem[\protect\citeauthoryear{Corbelli, Verley, Elmegreen et~al.,}{Corbelli
  et~al.}{2009}]{Corbelli:2009p781}
Corbelli E.,  Verley S.,  Elmegreen B.~G.,    et~al., 2009, A\&A, 495, 479

\bibitem[\protect\citeauthoryear{Dale, Cohen, Johnson et~al.,}{Dale
  et~al.}{2009}]{Dale:2009p765}
Dale D.~A.,  Cohen S.~A.,  Johnson L.~C.,    et~al., 2009, ApJ, 703, 517

\bibitem[\protect\citeauthoryear{de Paz, Boissier, Madore et~al.,}{de~Paz
  et~al.}{2007}]{GildePaz:2007p692}
de Paz A.~G.,  Boissier S.,  Madore B.~F.,    et~al., 2007, ApJS, 173, 185

\bibitem[\protect\citeauthoryear{Dove, Shull \& Ferrara}{Dove
  et~al.}{2000}]{Dove:2000p753}
Dove J.~B.,  Shull J.~M.,    Ferrara A.,  2000, ApJ, 531, 846

\bibitem[\protect\citeauthoryear{Eldridge}{Eldridge}{2011}]{Eldridge:2011p755}
Eldridge J.~J.,  2011, arXiv, 1106, 4311

\bibitem[\protect\citeauthoryear{Eldridge \& Rela{\~n}o}{Eldridge \&
  Rela{\~n}o}{2011}]{Eldridge:2011p756}
Eldridge J.~J.,  Rela{\~n}o M.,  2011, MNRAS, 411, 235

\bibitem[\protect\citeauthoryear{Eldridge \& Stanway}{Eldridge \&
  Stanway}{2009}]{Eldridge:2009p778}
Eldridge J.~J.,  Stanway E.~R.,  2009, MNRAS, 400, 1019

\bibitem[\protect\citeauthoryear{Fitzpatrick, Ribas, Guinan
  et~al.,}{Fitzpatrick et~al.}{2002}]{Fitzpatrick:2002p741}
Fitzpatrick E.~L.,  Ribas I.,  Guinan E.~F.,    et~al., 2002, ApJ, 564, 260

\bibitem[\protect\citeauthoryear{Freedman, Wilson \& Madore}{Freedman
  et~al.}{1991}]{Freedman:1991p485}
Freedman W.~L.,  Wilson C.~D.,    Madore B.~F.,  1991, ApJ, 372, 455

\bibitem[\protect\citeauthoryear{Fumagalli, da Silva \& Krumholz}{Fumagalli
  et~al.}{2011}]{Fumagalli:2011p771}
Fumagalli M.,  da Silva R.~L.,    Krumholz M.~R.,  2011, arXiv, astro-ph.CO

\bibitem[\protect\citeauthoryear{Gerola, Seiden \& Schulman}{Gerola
  et~al.}{1980}]{Gerola:1980p780}
Gerola H.,  Seiden P.~E.,    Schulman L.~S.,  1980, ApJ, 242, 517

\bibitem[\protect\citeauthoryear{Giammanco, Beckman \& Cedr{\'e}s}{Giammanco
  et~al.}{2005}]{Giammanco:2005p625}
Giammanco C.,  Beckman J.~E.,    Cedr{\'e}s B.,  2005, A{\&}A, 438, 599

\bibitem[\protect\citeauthoryear{Giammanco, Beckman, Zurita et~al.,}{Giammanco
  et~al.}{2004}]{Giammanco:2004p626}
Giammanco C.,  Beckman J.~E.,  Zurita A.,    et~al., 2004, A{\&}A, 424, 877

\bibitem[\protect\citeauthoryear{Gordon, Clayton, Witt et~al.,}{Gordon
  et~al.}{2000}]{Gordon:2000p742}
Gordon K.~D.,  Clayton G.~C.,  Witt A.~N.,    et~al., 2000, ApJ, 533, 236

\bibitem[\protect\citeauthoryear{Grossi, Corbelli, Giovanardi et~al.,}{Grossi
  et~al.}{2010}]{Grossi:2010p754}
Grossi M.,  Corbelli E.,  Giovanardi C.,    et~al., 2010, A\&A, 521, 41

\bibitem[\protect\citeauthoryear{Hanish, Oey, Rigby et~al.,}{Hanish
  et~al.}{2010}]{Hanish:2010p750}
Hanish D.~J.,  Oey M.~S.,  Rigby J.~R.,    et~al., 2010, ApJ, 725, 2029

\bibitem[\protect\citeauthoryear{Hirashita, Buat \& Inoue}{Hirashita
  et~al.}{2003}]{Hirashita:2003p768}
Hirashita H.,  Buat V.,    Inoue A.~K.,  2003, A\&A, 410, 83

\bibitem[\protect\citeauthoryear{Hoopes \& Walterbos}{Hoopes \&
  Walterbos}{2000}]{Hoopes:2000p594}
Hoopes C.~G.,  Walterbos R. A.~M.,  2000, ApJ, 541, 597

\bibitem[\protect\citeauthoryear{Hunter, Elmegreen \& Ludka}{Hunter
  et~al.}{2010}]{Hunter:2010p713}
Hunter D.~A.,  Elmegreen B.~G.,    Ludka B.~C.,  2010, AJ, 139, 447

\bibitem[\protect\citeauthoryear{Iglesias-P{\'a}ramo, Boselli, Gavazzi
  et~al.,}{Iglesias-P{\'a}ramo et~al.}{2004}]{IglesiasParamo:2004p769}
Iglesias-P{\'a}ramo J.,  Boselli A.,  Gavazzi G.,    et~al., 2004, A\&A, 421,
  887

\bibitem[\protect\citeauthoryear{Kennicutt}{Kennicutt}{1998}]{Kennicutt:1998p6%
11}
Kennicutt R.~C.,  1998, ARA\&A, 36, 189

\bibitem[\protect\citeauthoryear{Kennicutt, Calzetti, Walter et~al.,}{Kennicutt
  et~al.}{2007}]{Kennicutt:2007p488}
Kennicutt R.~C.,  Calzetti D.,  Walter F.,    et~al., 2007, ApJ, 671, 333

\bibitem[\protect\citeauthoryear{Kennicutt \& Hodge}{Kennicutt \&
  Hodge}{1986}]{Kennicutt:1986p772}
Kennicutt R.~C.,  Hodge P.~W.,  1986, ApJ, 306, 130

\bibitem[\protect\citeauthoryear{Kennicutt, Lee, Funes et~al.,}{Kennicutt
  et~al.}{2008}]{Kennicutt:2008p678}
Kennicutt R.~C.,  Lee J.~C.,  Funes S.~J.,    et~al., 2008, ApJS, 178, 247

\bibitem[\protect\citeauthoryear{Kramer, Buchbender, Xilouris et~al.,}{Kramer
  et~al.}{2010}]{Kramer:2010p688}
Kramer C.,  Buchbender C.,  Xilouris E.~M.,    et~al., 2010, A\&A, 518, L67

\bibitem[\protect\citeauthoryear{Lee, de Paz, Kennicutt et~al.,}{Lee
  et~al.}{2010}]{Lee:2010p708}
Lee J.~C.,  de Paz A.~G.,  Kennicutt R.~C.,    et~al., 2010, ApJS, 192, 6

\bibitem[\protect\citeauthoryear{Lee, de Paz, Kennicutt et~al.,}{Lee
  et~al.}{2011}]{Lee:2011p766}
Lee J.~C.,  de Paz A.~G.,  Kennicutt R.~C.,    et~al., 2011, The Astrophysical
  Journal Supplement, 192, 6

\bibitem[\protect\citeauthoryear{Lee, de Paz, Tremonti et~al.,}{Lee
  et~al.}{2009}]{Lee:2009p609}
Lee J.~C.,  de Paz A.~G.,  Tremonti C.,    et~al., 2009, ApJ, 706, 599

\bibitem[\protect\citeauthoryear{McKee \& Williams}{McKee \&
  Williams}{1997}]{McKee:1997p802}
McKee C.~F.,  Williams J.~P.,  1997, Astrophysical Journal v.476, 476, 144

\bibitem[\protect\citeauthoryear{Mcquinn, Skillman, Cannon et~al.,}{Mcquinn
  et~al.}{2009}]{Mcquinn:2009p709}
Mcquinn K. B.~W.,  Skillman E.~D.,  Cannon J.~M.,    et~al., 2009, ApJ, 695,
  561

\bibitem[\protect\citeauthoryear{Martin, Fanson, Schiminovich et~al.,}{Martin
  et~al.}{2005}]{Martin:2005p691}
Martin D.~C.,  Fanson J.,  Schiminovich D.,    et~al., 2005, ApJ, 619, L1

\bibitem[\protect\citeauthoryear{Massey, Olsen, Hodge et~al.,}{Massey
  et~al.}{2006}]{Massey:2006p517}
Massey P.,  Olsen K. A.~G.,  Hodge P.~W.,    et~al., 2006, AJ, 131, 2478

\bibitem[\protect\citeauthoryear{Melena, Elmegreen, Hunter et~al.,}{Melena
  et~al.}{2009}]{Melena:2009p775}
Melena N.~W.,  Elmegreen B.~G.,  Hunter D.~A.,    et~al., 2009, AJ, 138, 1203

\bibitem[\protect\citeauthoryear{Meurer, Wong, Kim et~al.,}{Meurer
  et~al.}{2009}]{Meurer:2009p608}
Meurer G.~R.,  Wong O.~I.,  Kim J.~H.,    et~al., 2009, ApJ, 695, 765

\bibitem[\protect\citeauthoryear{Oey \& Kennicutt}{Oey \&
  Kennicutt}{1997}]{Oey:1997p612}
Oey M.~S.,  Kennicutt R.~C.,  1997, MNRAS, 291, 827

\bibitem[\protect\citeauthoryear{Oey, Meurer, Yelda et~al.,}{Oey
  et~al.}{2007}]{Oey:2007p749}
Oey M.~S.,  Meurer G.~R.,  Yelda S.,    et~al., 2007, ApJ, 661, 801

\bibitem[\protect\citeauthoryear{Pflamm-Altenburg, Weidner \&
  Kroupa}{Pflamm-Altenburg et~al.}{2009}]{PflammAltenburg:2009p675}
Pflamm-Altenburg J.,  Weidner C.,    Kroupa P.,  2009, MNRAS, 395, 394

\bibitem[\protect\citeauthoryear{Rela{\~n}o \& Beckman}{Rela{\~n}o \&
  Beckman}{2005}]{Relano:2005p644}
Rela{\~n}o M.,  Beckman J.~E.,  2005, A{\&}A, 430, 911

\bibitem[\protect\citeauthoryear{Rela{\~n}o et~al.,}{Rela{\~n}o
  et~al.}{2012}]{Relano:2012}
Rela{\~n}o M.,  et~al., 2012, in preparation

\bibitem[\protect\citeauthoryear{Rela{\~n}o, Peimbert \& Beckman}{Rela{\~n}o
  et~al.}{2002}]{Relano:2002p97}
Rela{\~n}o M.,  Peimbert M.,    Beckman J.,  2002, ApJ, 564, 704

\bibitem[\protect\citeauthoryear{Sullivan, Treyer, Ellis et~al.,}{Sullivan
  et~al.}{2004}]{Sullivan:2004p707}
Sullivan M.,  Treyer M.~A.,  Ellis R.~S.,    et~al., 2004, MNRAS, 350, 21

\bibitem[\protect\citeauthoryear{van~den Bergh}{van~den
  Bergh}{2000}]{vandenBergh:2000p502}
van~den Bergh S.,  2000, The galaxies of the Local Group

\bibitem[\protect\citeauthoryear{Verley, Corbelli, Giovanardi et~al.,}{Verley
  et~al.}{2009}]{Verley:2009p573}
Verley S.,  Corbelli E.,  Giovanardi C.,    et~al., 2009, A{\&}A, 493, 453

\bibitem[\protect\citeauthoryear{Verley, Hunt, Corbelli et~al.,}{Verley
  et~al.}{2007}]{Verley:2007p574}
Verley S.,  Hunt L.~K.,  Corbelli E.,    et~al., 2007, A{\&}A, 476, 1161

\bibitem[\protect\citeauthoryear{Verley, Rela{\~n}o, Kramer et~al.,}{Verley
  et~al.}{2010}]{Verley:2010p687}
Verley S.,  Rela{\~n}o M.,  Kramer C.,    et~al., 2010, A\&A, 518, L68

\bibitem[\protect\citeauthoryear{Weidner \& Kroupa}{Weidner \&
  Kroupa}{2005}]{Weidner:2005p800}
Weidner C.,  Kroupa P.,  2005, ApJ, 625, 754

\bibitem[\protect\citeauthoryear{Weisz, Johnson, Johnson et~al.,}{Weisz
  et~al.}{2012}]{Weisz:2012p784}
Weisz D.~R.,  Johnson B.~D.,  Johnson L.~C.,    et~al., 2012, ApJ, 744, 44

\bibitem[\protect\citeauthoryear{Weisz, Skillman, Cannon et~al.,}{Weisz
  et~al.}{2008}]{Weisz:2008p723}
Weisz D.~R.,  Skillman E.~D.,  Cannon J.~M.,    et~al., 2008, ApJ, 689, 160

\bibitem[\protect\citeauthoryear{Whitmore, Chandar, Kim et~al.,}{Whitmore
  et~al.}{2011}]{Whitmore:2011p726}
Whitmore B.~C.,  Chandar R.,  Kim H.,    et~al., 2011, ApJ, 729, 78

\bibitem[\protect\citeauthoryear{Wood, Hill, Joung et~al.,}{Wood
  et~al.}{2010}]{Wood:2010p801}
Wood K.,  Hill A.~S.,  Joung M.~R.,    et~al., 2010, ApJ, 721, 1397

\bibitem[\protect\citeauthoryear{Zurita, Beckman, Rozas et~al.,}{Zurita
  et~al.}{2002}]{Zurita:2002p747}
Zurita A.,  Beckman J.~E.,  Rozas M.,    et~al., 2002, A\&A, 386, 801

\bibitem[\protect\citeauthoryear{Zurita, Rozas \& Beckman}{Zurita
  et~al.}{2000}]{Zurita:2000p748}
Zurita A.,  Rozas M.,    Beckman J.~E.,  2000, A\&A, 363, 9

\end{thebibliography}

\label{lastpage}

\end{document}